\documentclass[sigconf]{acmart}
\usepackage{enumitem}
\usepackage{multirow}
\AtBeginDocument{%
  }

\begin{document}

\title{Transforming User-Defined Criteria into Explainable Indicators with an Integrated LLM–AHP System}


\author{Geonwoo Bang}
\affiliation{%
  \institution{Sungkyunkwan University}
  \city{Seoul}
  \country{Republic of Korea}}
\email{g7199@g.skku.edu}

\author{Dongho Kim}
\affiliation{%
  \institution{Sungkyunkwan University}
  \city{Seoul}
  \country{Republic of Korea}}
\email{dpl9753@g.skku.edu}

\author{Moohong Min}
\authornote{Corresponding author}
\affiliation{%
  \institution{Sungkyunkwan University}
  \city{Seoul}
  \country{Republic of Korea}}
\email{iceo@skku.edu}

\renewcommand{\shortauthors}{Bang et al.}

\begin{abstract}
    Evaluating complex texts across domains requires converting user defined criteria into quantitative, explainable indicators, which is a persistent challenge in search and recommendation systems. Single-prompt LLM evaluations suffer from complexity and latency issues, while criterion-specific decomposition approaches rely on naive averaging or opaque black-box aggregation methods. We present an interpretable aggregation framework combining LLM scoring with the Analytic Hierarchy Process (AHP). Our method generates criterion-specific scores via LLM-as-judge, measures discriminative power using Jensen–Shannon distance, and derives statistically grounded weights through AHP pairwise comparison matrices. Experiments on Amazon review quality assessment and depression related text scoring demonstrate that our approach achieves high explainability and operational efficiency while maintaining comparable predictive power, making it suitable for real-time latency sensitive web services.

\end{abstract}




\maketitle

\section{Introduction}
Text quality evaluation is crucial in various web and data mining applications, such as review recommendation \cite{moghaddam2011reviewrec, zheng2017deepconn, qu2021review, chen2022argument}, content moderation \cite{qiao2024scaling, huang2025content}, and survey analysis \cite{mcgillivray2020extracting, mellon2024do}. Traditional methods often rely on human annotations or heuristic metrics (e.g., readability scores), but these are costly and domain-specific. Recent advances in large language models (LLMs) have enabled automated evaluation through frameworks like G-Eval \cite{liu2023geval}, which prompt LLMs to score texts on user-defined criteria using Likert scales \cite{likert1932technique}.

However, directly using LLMs for comprehensive quality assessment poses challenges. Small-scale models struggle with complex tasks due to limited capacity \cite{wang2025survey, diehlmartinez2024tending}, while large models incur high computational costs \cite{fernandez-etal-2025-energy, stojkovic2024towards}. These constraints often force practitioners to limit both model size and the number of evaluation runs, which exacerbates score instability and bias. To address this, we advocate decomposing evaluation into lightweight, per-criterion assessments with small models, then aggregating them into a unified score via a robust integration framework. A natural choice for such an aggregator might be linear regression, fitting weights to predict observed signals like vote counts from criteria scores. However, LLM-generated scores often suffer from central tendency bias, where outputs cluster around the middle of the scale with few extreme values \cite{rupprecht2025prompt}. This compression distorts regression weights, sometimes undervaluing criteria that are actually highly discriminative \cite{lee2025systematic}. While normalization may seem like a fix, it performs poorly on discrete Likert-scale outputs and cannot restore variance absent in the original data. Such imbalance not only reduces predictive accuracy but also undermines the aggregator’s interpretability, making it hard for users to understand why certain criteria dominate the final score \cite{bojic2025comparing, li2024llms}.

To overcome these limitations, we propose UniScore, an interpretable framework that combines multiple criterion-specific scores from LLM evaluations into a single robust indicator, even when using small models or limited inference runs. UniScore measures each criterion’s discriminative power using the Jensen–Shannon distance (JSD) \cite{endres2003new} and assigns weights through the Analytic Hierarchy Process (AHP)~\cite{Saaty1980}. By relying on relative pairwise comparisons rather than absolute scale values, UniScore mitigates bias from skewed or clustered LLM outputs, producing interpretable signed weights: criteria with higher discriminative power receive proportionally larger magnitudes, while the sign reflects directional quality (positive for beneficial associations, negative otherwise). This design enables transparent and scalable aggregation, maintaining reliability in low-cost evaluation pipelines.

Our main contributions are:

\begin{itemize}[leftmargin=20pt, itemsep=2pt]
\item A novel JSD–AHP aggregation method that corrects scale bias and instability in LLM-generated scores, making it suitable for small-model and low-repetition settings.
\item Empirical validation on multiple datasets demonstrating superior correlation with external signals over other baselines.
\item Evidence of efficiency and interpretability, supporting deployment in real-time web applications.
\end{itemize}

The rest of the paper is organized as follows: Section~\ref{sec:related works} reviews related work, Section~\ref{sec:preliminaries} introduces preliminaries, Section~\ref{sec:methods} describes the proposed UniScore method, Section~\ref{sec:experiments} presents experimental results, Section~\ref{sec:discussion} discusses implications and limitations, and Section~\ref{sec:conclusion} concludes the paper.

\section{Related Works}
\label{sec:related works}

In this section, we review prior work relevant to predictive text quality evaluation, with a focus on methods applicable to real-time web applications.

\subsection{Predictive Text Scoring and Challenges}
Predictive text quality evaluation is central to web services such as review recommendations~\cite{moghaddam2011reviewrec, zheng2017deepconn, qu2021review, chen2022argument}, content moderation~\cite{qiao2024scaling, huang2025content}, and survey analysis~\cite{mcgillivray2020extracting, mellon2024do}. Traditional heuristic-based metrics like Flesch-Kincaid readability scores and lexical complexity \cite{Kincaid1975} enable real-time evaluation but fail to capture deep semantic qualities such as logical coherence, expertise, or persuasiveness \cite{louis2013corpus}. 

Conversely, user votes (e.g., \textit{helpful} ratings) provide valuable quality assessments \cite{zhang2014helpfulness, singh2017predicting} but are post-hoc metrics unavailable for new content. These limitations have motivated the exploration of automated evaluators capable of delivering semantically rich assessments, leading to growing interest in LLM-based approaches.

\subsection{Limitations of LLM-based Evaluators}
To bridge the gap between shallow heuristics and delayed user-vote signals, recent advances in LLMs have enabled automated evaluation through frameworks like G-Eval \cite{liu2023geval}, which prompt LLMs to score texts on user-defined criteria using Likert scales \cite{likert1932technique}. 

However, LLM evaluation faces a Performance vs. Cost Dilemma \cite{shashidhar2023democratizing}: while massive models like GPT-5 \cite{openai2025gpt5} and Claude Sonnet 4 \cite{anthropic2025claude4} exhibit high accuracy, their computational cost and slow inference make them unsuitable for real-time applications \cite{zhou2024survey}. Small-scale models are cost-effective but may lack reliable evaluation capabilities \cite{zheng2023judging}. Additionally, LLMs are sensitive to prompt variations, hindering consistency \cite{sclar2024quantifying}. Particularly in real-time environments like review recommendation systems, where a review must be ranked immediately upon submission, there is a strict latency constraint that requires the evaluation to be completed within a few milliseconds to a few seconds \cite{barros2025solving}. Under such constraints, approaches that require extensive feature extraction or multi-turn LLM queries are impractical, as even minor delays can degrade user experience or reduce the freshness of ranking outputs. 

Recent research addresses these limitations by decomposing overall quality into clear sub-criteria, thereby reducing the cognitive and computational burden on LLMs and improving score consistency. For instance, FLASK \cite{ye2023flask} decomposes coarse-level evaluation into fine-grained skill dimensions, enabling detailed diagnosis of a model across multiple capabilities. While such decomposition improves reliability, existing methods remain primarily diagnostic in nature and do not yield a unified, operational score, limiting their direct applicability in downstream systems.

\subsection{Aggregating Multi-Criteria Scores}
Even if we obtain individual scores for multiple criteria (e.g., readability, expertise, originality) via an LLM, the core question remains: how should these scores be aggregated into a single, comprehensive quality metric that is both predictive and interpretable?

Naïve approaches such as simple averaging assume equal importance for all criteria, which overlooks the varying discriminative power of individual dimensions. Linear regression can estimate weights from data, but recent studies show that LLM-generated scores suffer from skewed and compressed distributions, often with central tendency bias~\cite{stureborg2024large,wang2023large}. Such compression, especially under cost or latency constraints with lightweight models, reduces usable variance and amplifies noise. Liddell et al.~\cite{Liddell2018} note that treating Likert-scale outputs as continuous variables in regression violates key statistical assumptions, leading to unstable or misleading coefficients. Even averaging multiple runs cannot recover variance that was absent in the first place.

Some frameworks, such as HD-Eval \cite{liu2024hdeval}, make valuable contributions by incorporating human preference data to train aggregation models that combine decomposed evaluation scores. These methods have proven effective for their intended purpose as offline benchmarks. However, they are not primarily designed for low-latency, lightweight scenarios, and their aggregation models (e.g., regression, random forests, neural networks) can face challenges with interpretability, scaling biases, and label type or distribution issues, as well as retraining overhead in dynamic environments.

These limitations highlight the need for an alternative approach grounded in Multi-Criteria Decision Analysis (MCDA). In this study, we adopt the Analytic Hierarchy Process (AHP) \cite{Saaty1980} to derive relative weights from the statistical discriminative power of each criterion, enabling consistent, interpretable aggregation without assuming uniform scaling or continuous score distributions. 

\medskip
This study systematically integrates validated principles from Information Theory and MCDA to address the score aggregation challenge.
Recent research has explored leveraging LLMs directly within AHP's pairwise comparison stage. Lu et al. (2024) \cite{lu2024ahp} present evaluation criteria to an LLM and explicitly ask "How much more important is criterion A than criterion B?" to populate the pairwise comparison matrix, treating the LLM as an automated expert relying on qualitative reasoning.

UniScore takes a different approach. Rather than relying on LLM's subjective judgments, our framework treats the LLM as a scalable scorer and derives weights from empirical data. First, we employ JSD to measure each criterion's discriminative power by comparing score distributions between signal-based groups defined by external quality signals. JSD is a symmetric, bounded divergence metric focusing on distribution shapes rather than absolute values, enabling detection of meaningful differences even when scores are narrowly concentrated due to scaling or central tendency bias, yielding objective, data-driven discriminativeness measures. Second, we apply AHP to convert these discriminative power measurements into final weights by computing differences in JSD between criterion pairs and transforming them into pairwise importance ratios. AHP derives relative weights based on dominance rather than absolute magnitude, making the resulting weights consistent, interpretable, and robust while avoiding distributional assumptions. Each weight's reasoning can be directly traced through explicit pairwise comparisons of discriminative power.

In summary, while previous studies rely on LLM's own judgment, UniScore's strategy is data-driven. By quantifying distributional differences via JSD and transforming them into relative weights through AHP, UniScore produces robust, interpretable, and objective aggregation particularly well-suited for real-time applications requiring low-latency evaluation combined with transparent decision-making.

\section{Preliminaries}
\label{sec:preliminaries}

UniScore relies on two key components: measuring discriminative power between groups using information-theoretic distance, and deriving interpretable weights through structured decision-making. We outline the essential concepts below.

\subsection{Jensen–Shannon Distance}
\label{sec:jsd_pre}

The Jensen–Shannon divergence (JSDiv) is a symmetric, finite divergence measure for quantifying differences between probability distributions, defined as a symmetrized version of the Kullback–Leibler divergence \cite{kullback1951information, lin1991divergence}. Since JSDiv does not satisfy the triangle inequality, we use its square root $d=\sqrt{\text{JSDiv}}$, which forms a true metric satisfying the triangle inequality \cite{endres2003new, nielsen2020generalization}.

We compute JSDiv between score distributions of two groups for each criterion, then use $d=\sqrt{\text{JSDiv}}$ as the metric distance representing discriminative power. This approach leverages metric space properties that allow both addition and subtraction of distances, providing a mathematically rigorous and interpretable measure.

JSDiv is particularly suitable for this application as it produces bounded values in [0,1], ensuring consistent weighting in the AHP framework. Importantly, due to the statistical characteristics of LLM responses, even small absolute differences can result in completely separable distributions, and JSDiv guarantees a maximum value of 1 for such cases, providing reliable discrimination detection.

\subsection{Analytic Hierarchy Process}
\label{sec:ahp_pre}

The Analytic Hierarchy Process (AHP) \cite{Saaty1980} is a multi criteria decision making method that quantifies relative criterion importance through pairwise comparisons. The method structures decisions as hierarchies with goals, criteria, and alternatives, then uses pairwise comparisons expressed as positive ratios $a_{ij}$ indicating how many times criterion $i$ is more important than criterion $j$.

The pairwise comparison matrix $A=[a_{ij}]$ is positive and reciprocal with $a_{ij}=1/a_{ji}$ and $a_{ii}=1$. Criterion weights are obtained as the normalized principal right eigenvector:

\begin{equation}
A\,\mathbf{w}=\lambda_{\max}\,\mathbf{w},\qquad
\hat{\mathbf{w}}=\frac{\mathbf{w}}{\sum_{k=1}^{n} w_k}
\label{eq:eig-weights}
\end{equation}

where $\lambda_{\max}$ is the largest eigenvalue and $\mathbf w$ is the corresponding eigenvector.

Consistency is measured by the consistency ratio (CR), which should not exceed 0.1 for reliable results \cite{saaty1990how}. When multiple experts provide judgments, AHP aggregates individual pairwise entries using geometric means \cite{aczel1983procedures}.

In our framework, pairwise information is generated mechanically from quantitative Jensen–Shannon distances rather than human expert judgments. This construction enforces consistency by design and yields data-driven comparison matrices compatible with the eigenvector method. The AHP framework explicitly allows relative scales derived from data mapped to standard ratio scales \cite{saaty1977scaling}, justifying our distance-derived entries. The resulting weights are ratio-scale linear coefficients that are uniquely determined and reproducible, providing mathematical rigor and reliability for score aggregation.

\section{Methods}
\label{sec:methods}

\begin{figure*}[t]
 \centering
 \includegraphics[width=\textwidth]{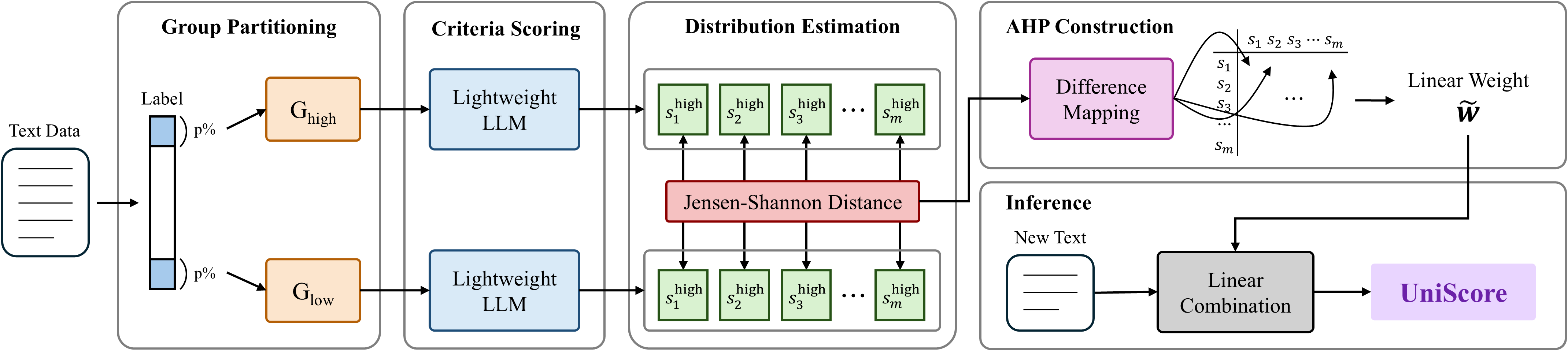} 
 \caption{The overall architecture of the UniScore framework on Continuous Signals.}
 \label{fig:framework_overview}
\end{figure*}

In this section, we introduce our proposed \textbf{UniScore}(Unified Scoring Framework). First, we describe the process of obtaining individual scores for user-defined criteria using LLM-based evaluation. Then, we explain how these criterion-level scores are integrated into a single interpretable scoring formula through AHP-based weight estimation. The overall architecture of this framework is illustrated in Figure~\ref{fig:framework_overview}.

\subsection{Group Partitioning}
\label{sec:group_partitioning}

Our framework operates under a supervised setting, requiring an observed signal that reflects some existing evaluation of each text sample. This signal can originate from arbitrary sources, such as star ratings, clinical depression diagnoses, vote counts, or visitor statistics, depending on the task domain. The observed signal is used solely for partitioning purposes and does not need to be aligned with any of the user-defined evaluation criteria introduced later.

Each text sample $x_i$ is represented as:
\begin{equation}
    x_i = (t_i, s_i)
\end{equation}

\noindent where:

\begin{itemize}[leftmargin=20pt, itemsep=2pt]
    \item $t_i$: textual content (e.g., review body, essay, survey response)
    \item $s_i$: observed signal for group partitioning
\end{itemize}

\medskip
\noindent The observed signal $s_i$ may be:
\begin{itemize}[leftmargin=20pt, itemsep=2pt]
    \item \textbf{Discrete}: e.g., depressed vs. non-depressed, pass vs. fail
    \item \textbf{Continuous}: e.g., vote count, numerical score, time spent
\end{itemize}

\noindent The partitioning rules can be divided into two general cases.
\medskip

\textbf{Discrete signals.}
If the dataset is already separated into two distinct categories, the groups are defined directly as:
\begin{equation}
G_{\text{low}} = \{ x_i \mid s_i = 0 \}, \quad G_{\text{high}} = \{ x_i \mid s_i = 1 \}
\end{equation}
To ensure computational efficiency and statistical balance, we provide two sampling strategies:
\medskip

\emph{Size balancing.} When there is a significant imbalance in group sizes, random sampling is performed from the larger group to match the size of the smaller one, ensuring $|G_{\text{low}}| = |G_{\text{high}}|$.
\medskip

\emph{Computational efficiency.} For large datasets, the user may specify a target sample size $n_{\text{target}}$, in which case exactly $n_{\text{target}}$ samples are randomly drawn from each group:
\begin{equation}
|G_{\text{low}}| = |G_{\text{high}}| = n_{\text{target}}
\end{equation}
This enables efficient computation without processing the entire dataset while maintaining group balance.

\medskip
\textbf{Continuous signals.}  
For continuous-valued signals, the user specifies a percentile threshold $p$ (e.g., $5\%$), which determines the upper and lower boundaries:
\begin{equation}
    \tau_{\text{low}} = Q_p(S), \quad \tau_{\text{high}} = Q_{1-p}(S)
\end{equation}
where $Q_p(S)$ denotes the $p$-th percentile of the signal set $S = \{ s_i \}$.  
The groups are then defined as:
\begin{equation}
G_{\text{low}} = \{ x_i \mid s_i \le \tau_{\text{low}} \}, \quad
G_{\text{high}} = \{ x_i \mid s_i \ge \tau_{\text{high}} \}
\end{equation}
If multiple samples have the same value at the percentile boundary, samples with values exceeding the boundary are included first, followed by random selection from the boundary-value samples to reach the exact percentile.

Through this process, the final two comparison groups $G_{\text{low}}$, $G_{\text{high}}$ are obtained, which serve as the input for the LLM based scoring procedure described in Section~\ref{sec:scoring}.

\subsection{Criteria Scoring}
\label{sec:scoring}

In UniScore, we adopt an LLM-based scoring procedure to produce per-criterion scores. The user first specifies a set of criteria
\begin{equation}
\mathcal{C} = \{ c_1, c_2, \dots, c_m \}
\end{equation}
where each $c_k$ denotes a textual property and is not restricted to a particular domain. For instance, in a product-review recommendation service, the designer may choose to evaluate overall sentiment ($c_1$), the reviewer's domain-specific expertise ($c_2$), the specificity of description with concrete details ($c_3$), and the consistency between the star rating and the review content ($c_4$). For each criterion $c_k$, the user writes a criteria prompt $P_k$ that instructs the model to return a 1--5 Likert score.
\medskip

\noindent Each LLM-as-judge prompt $P_k$ must include the following elements:

\begin{enumerate}[leftmargin=20pt, itemsep=2pt]
\item a clear \emph{definition} of the criterion;
\item concise \emph{guidelines} for assessment;
\item a \emph{scale description} specifying the meaning of scores 1--5; and
\item an \emph{output specification} requiring JSON-only output of the form \verb|{"score": N}|.
\end{enumerate}

The prompts $\{P_k\}$ are applied to the samples in the two groups ${G}_{\text{low}}$ and ${G}_{\text{high}}$ constructed in Section~\ref{sec:group_partitioning}. For each sample $x_i$ and criterion $c_k$, the LLM produces a discrete score
\begin{equation}
s_{ik} = \text{LLM}(P_k, x_i) \in \{1,2,3,4,5\}.
\end{equation}

Considering real-time web deployment, we employ \emph{lightweight} LLMs to reduce GPU memory usage and latency, enabling responsive scoring for hundreds to thousands of texts under practical resource constraints. To prioritize inference speed over stochastic variability, we set the temperature=0 for deterministic outputs, unlike the other LLM evaluation methods which use multiple runs for averaging; this may slightly reduce stability but ensures faster processing. To handle potential invalid outputs, such as non-JSON responses, we implement a retry mechanism up to three attempts or fallback to a neutral score of 3 for robustness.

For cost efficiency, we do not score the entire dataset. Instead, scoring is applied only to the subset extracted in Section~\ref{sec:group_partitioning}. The resulting per-criterion score vectors are:
\begin{equation}
\mathbf{s}_k^{\mathrm{low}} = \big[s_{ik}\big]_{i \in \mathcal{I}_{\mathrm{low}}}, \quad
\mathbf{s}_k^{\mathrm{high}} = \big[s_{ik}\big]_{i \in \mathcal{I}_{\mathrm{high}}}
\end{equation}

where $\mathcal{I}_{\mathrm{low}}$ and $\mathcal{I}_{\mathrm{high}}$ denote the index sets of samples in the low and high groups, respectively.

For purely quantitative criteria (e.g., word count, Flesch-Kincaid index), we bypass LLM scoring and directly scale the raw measurement to a 1--5 scale score. Users can apply various techniques based on the task requirements. For instance, one common scaling approach is:

\begin{equation}
z_i = \frac{r_i - \mu}{\sigma},\quad
\tilde z_i = \max\{1, \min\{5, z_i \cdot \sigma_{\text{scale}} + 3\}\}, \quad
s_{ik} = \tilde z_i
\label{eq:scale}
\end{equation}

\noindent where $r_i$ is the raw measurement; $\mu$ and $\sigma$ are the mean and standard deviation computed on the scored subset; $\sigma_{\text{scale}}$ is a parameter to map the standardized scores to the 1--5 range. This approach handles outliers by clipping extreme values while preserving the common 1--5 scale. In general, users can define various continuous variables tailored to their domain and apply appropriate scaling methods, such as the one above or alternatives like min-max normalization, to produce scores in [1,5]. The handling of such continuous scores is also addressed in Section~\ref{sec:jsd}.

\subsection{Distribution Estimation}
\label{sec:jsd}
To determine the relative importance of each criterion, we need to quantify how well each criterion discriminates between the high and low quality groups. This is achieved by comparing the score distributions of each criterion across the two groups using Jensen-Shannon distance.

Given the group-wise score matrices from Section~\ref{sec:scoring}, $\mathbf{S}^{\mathrm{low}}$ and $\mathbf{S}^{\mathrm{high}}$, we estimate, for each criterion $c_k$, the empirical probability mass functions (PMFs) over the Likert levels $\mathcal{L}=\{1,2,3,4,5\}$ for the low and high groups. For each criterion $k$ and Likert-scale value $v \in \mathcal{L}$, we define
\begin{equation}
n^{\mathrm{low}}_{k}(v) = \sum_{i \in G_{\mathrm{low}}} \mathbf{1}[s_{ik} = v],\quad
n^{\mathrm{high}}_{k}(v) = \sum_{i \in G_{\mathrm{high}}} \mathbf{1}[s_{ik} = v].
\end{equation}
These counts represent the score distributions for the two groups under criterion $k$.

With Laplace smoothing $\varepsilon>0$ (we use $\varepsilon=10^{-6}$) to avoid zero probabilities, the smoothed PMFs are
\begin{equation}
P_k^{\mathrm{low}}(v)=\frac{n^{\mathrm{low}}_{k}(v)+\varepsilon}{\sum_{u\in\mathcal{L}} \big(n^{\mathrm{low}}_{k}(u)+\varepsilon\big)}
\label{eq:pmfp}
\end{equation}
\begin{equation}
Q_k^{\mathrm{high}}(v)=\frac{n^{\mathrm{high}}_{k}(v)+\varepsilon}{\sum_{u\in\mathcal{L}} \big(n^{\mathrm{high}}_{k}(u)+\varepsilon\big)}.
\label{eq:pmfq}
\end{equation}
We quantify the discriminativeness of criterion $c_k$ via the Jensen-Shannon \emph{distance} (base 2):
\begin{equation}
d_k = \sqrt{\,\mathrm{JSDiv}\big(P_k^{\mathrm{low}}\!\parallel\!Q_k^{\mathrm{high}}\big)\,} \in [0,1],
\label{eq:jsd}
\end{equation}
where $\mathrm{JSDiv}$ is the Jensen-Shannon divergence with mixture $M_k=\tfrac{1}{2}(P_k^{\mathrm{low}}+Q_k^{\mathrm{high}})$.

For continuous (non-Likert) scores, we histogram both groups using common bin edges determined by Sturges' formula $\lceil \log_2(n) + 1 \rceil$ bins to ensure data-driven discretization. Users can adjust the binning method if desired, such as for specific data characteristics. We then apply Laplace smoothing and normalize to obtain PMFs $P_k^{\mathrm{low}}$ and $Q_k^{\mathrm{high}}$, and compute the Jensen-Shannon distance $d_k$ as in Eq.~\eqref{eq:jsd}.

To enable proper directional weighting in the final scoring formula (Section~\ref{sec:ahp}), we determine the direction from sample means.
\begin{equation}
\bar{s}_k^{\mathrm{low}}=\frac{1}{|G_{\mathrm{low}}|}\sum_{i\in G_{\mathrm{low}}} s_{ik},
\quad
\bar{s}_k^{\mathrm{high}}=\frac{1}{|G_{\mathrm{high}}|}\sum_{i\in G_{\mathrm{high}}} s_{ik}.
\end{equation}
The direction indicator is:
\begin{equation}
sign_k=\operatorname{sign}\!\big(\bar{s}_k^{\mathrm{high}}-\bar{s}_k^{\mathrm{low}}\big)
\end{equation}.

\subsection{AHP Construction via Difference Mapping}
\label{sec:ahp}
To transform the criterion discriminativeness values into meaningful weights for the final indicator, we employ the AHP, a well-established multi-criteria decision-making framework. We employ the AHP as a systematic framework for deriving criterion weights, leveraging pairwise comparisons to capture relative importance and ensuring mathematical consistency via the principal eigenvector method.

Using the discriminativeness values $d_k \in [0, 1]$ defined in the previous section, we construct AHP pairwise comparisons based on their differences, mapping each difference to a positive ratio scale to ensure compatibility with the eigenvector method. This approach does not enforce the perfect transitivity assumed in ratio-scale AHP, which is an intentional design choice aimed at preventing the distortion or flattening of local variations in the data that can result from enforcing transitivity.

Compared to traditional ratio-based alternatives, the difference-based comparison offers several advantages:
\begin{enumerate}[leftmargin=15pt, itemsep=2pt]
    \item it avoids numerical instabilities and ratio inflation when some $d_j$ values approach zero,
    \item unlike exponential function applied to satisfy the ratio scale, it exhibits lower sensitivity to outlier values,
    \item it maintains an intuitive linear interpretation in which larger differences correspond to stronger relative preferences, making the results easier to explain to non-technical stakeholders.
\end{enumerate}
\medskip

For two criteria $c_i$ and $c_j$, we define the difference
\begin{equation}
\Delta_{ij} = d_i - d_j \in [-1,\,1].
\end{equation}
The pairwise comparison entry is then constructed as
\begin{equation}
a_{ij} = 
\begin{cases}
1 + 8\,\Delta_{ij}, & \text{if } \Delta_{ij} \ge 0,\\[4pt]
\dfrac{1}{\,1 + 8|\Delta_{ij}|\,}, & \text{if } \Delta_{ij} < 0,
\end{cases}
\qquad a_{ji} = 1/a_{ij}, \quad a_{ii} = 1.
\end{equation}
This mapping ensures several desirable properties. First, reciprocity is preserved exactly by construction: $a_{ij} \cdot a_{ji} = 1$ for all $i \neq j$. Second, the result adheres to Saaty's recommended range~\cite{Saaty1980} $a_{ij} \in [1/9, 9]$. When $\Delta_{ij} \geq 0$, the linear transformation $1 + 8\Delta_{ij}$ maps $[0, 1]$ to $[1, 9]$; the coefficient 8 is chosen to achieve this scaling, as it corresponds to the slope $(9-1)/1 = 8$, ensuring that the maximum possible difference $\Delta_{ij} = 1$ maps to the upper bound of 9. When $\Delta_{ij} < 0$, the reciprocal form $\frac{1}{1 + 8|\Delta_{ij}|}$ maps $(0, 1]$ to $[1/9, 1)$, ensuring the full valid range. Third, we adopt linear mapping $1 + 8\Delta_{ij}$ rather than nonlinear alternatives (e.g., $1 + 8\Delta_{ij}^2$, $\Delta_{ij}^9$) to ensure interpretational stability: equal differences in discriminativeness $|\Delta_{ij}|$ translate to proportional differences in pairwise preference intensity, facilitating consistent weight interpretation across different datasets and criterion combinations. Fourth, the mapping remains interpretable: when $d_i > d_j$, criterion $c_i$ is deemed more important than $c_j$ with intensity proportional to $|\Delta_{ij}|$.

Let $\mathbf{A} = [a_{ij}] \in \mathbb{R}^{m \times m}$ denote the resulting pairwise comparison matrix. Following standard AHP procedure, we obtain the criterion weights by computing and normalizing the principal eigenvector of $\mathbf{A}$:
\begin{equation}
A\,\mathbf{v} = \lambda_{\max}\,\mathbf{v}, 
\quad 
\mathbf{w} = \frac{\mathbf{v}}{\mathbf{1}^\top \mathbf{v}} = (w_1,\dots,w_m),
\quad 
\sum_{k=1}^m w_k = 1 .
\end{equation}
where $\lambda_{\max}$ is the largest eigenvalue and $\mathbf{v}$ is the corresponding eigenvector with positive entries.

 While difference-based approaches may raise concerns about potential transitivity violations, it is preserved to some extent due to the linear scaling from $\Delta_{ij}$. Furthermore, the advantages of this method outweigh these theoretical limitations compared to exponential mappings which satisfy transitivity perfectly but suffer from extreme sensitivity to outlier discriminativeness values. Consequently, the pairwise comparison matrices generally achieve strong logical consistency, as reflected in standard measures such as the Consistency Ratio (CR), with experimental results showing CR values well below 0.1 in the vast majority of cases, with additional details provided in Section~\ref{sec:consistency_exp}.

The final signed weights are obtained by incorporating the directional information:
\begin{equation}
\tilde{w}_k = sign_k \cdot w_k,
\end{equation}
where $w_k$ is the positive AHP-derived weight. This ensures that criteria with higher scores in the high-quality group receive positive weights, while criteria with higher scores in the low-quality group receive negative weights in the final scoring formula.

\subsection{Inference}
\label{sec:inference}

Once the signed weights $\tilde{\mathbf{w}} = (\tilde w_1, \dots, \tilde w_m)$ are estimated from the partitioned groups as described in the previous sections, UniScore can be efficiently applied to new texts in an inference phase. This phase does not require re-partitioning the data or recomputing distributions and weights, making it suitable for real-time deployment.

For a new text $x^*$, the per-criterion scores $\hat s_k(x^*) \in [1, 5]$ are obtained using the same LLM scoring prompts $P_k$ from Section~\ref{sec:scoring} for qualitative criteria or the user-defined scaling methods for quantitative criteria. The final UniScore is then computed as the weighted linear combination:
\begin{equation}
\mathrm{UniScore}(x^*) = \sum_{k=1}^{m} \tilde w_k \, \hat s_k(x^*) = \tilde{\mathbf{w}}^{\top} \hat{\mathbf{s}}(x^*),
\end{equation}
where $\hat{\mathbf{s}}(x^*) = (\hat s_1(x^*), \dots, \hat s_m(x^*))$.
The time complexity of inference is O(m) per text, primarily dominated by the m LLM calls (or quantitative computations), which is mitigated by the lightweight and deterministic setup detailed in Section~\ref{sec:scoring}. This formulation provides an interpretable quality score that reflects both the discriminative power of each criterion (via $|w_k|$) and its directional relationship with overall quality (via $sign_k$), while enabling fast inference through lightweight LLMs and deterministic processing as detailed in Section~\ref{sec:scoring}.


\section{Experiments}
\label{sec:experiments}

To validate the effectiveness and generalizability of UniScore, we design a comprehensive set of experiments. Our primary objective is to demonstrate that UniScore generates a more predictive and discriminative quality score compared to several intuitive baselines. We aim to answer three key research questions: 
(1) Does UniScore produce scores that correlate more strongly with ground-truth signals (e.g., user votes, expert labels) than baseline methods? 
(2) Is UniScore efficient enough to be deployed in real-world web services, delivering high-quality evaluations under limited computational resources?
(3) Are the weights derived by UniScore’s AHP-based aggregation process reliable and consistent with domain knowledge, thereby enhancing interpretability?

We conduct experiments with diverse baseline methods to ensure a rigorous and multifaceted evaluation.

\subsection{Datasets}
\label{subsec:datasets}

To demonstrate the generalizability of UniScore, we select publicly available datasets with diverse characteristics, spanning both continuous and discrete ground-truth signals. We particularly focus on datasets composed of naturally occurring human-generated texts to evaluate real-world web deployment scenarios where UniScore would be applied to authentic user-generated content such as reviews, posts, and responses. This diversity ensures our evaluation is not tailored to a single domain or signal type. All datasets are split into training and testing sets using an 80/20 ratio to ensure fair and consistent evaluation.
\medskip

\noindent
\textbf{Amazon Reviews.} This dataset provides user-written product reviews, where the number of \textit{helpful votes} serves as a continuous signal of text quality. We focus on the \textit{Software} category, which contains highly technical and information-dense reviews with distinct linguistic characteristics suitable for evaluating domain expertise and specificity. The data are sourced from the UCSD Amazon Review Dataset \cite{Ni2019RecsysJust}.

\medskip
\noindent
\textbf{RoSE XSum.} This dataset contains system-generated summaries with human-annotated ACU (Atomic Content Unit) scores as continuous signals, representing the proportion of reference information preserved in summaries~\cite{liu-etal-2023-towards-interpretable}.

\medskip
\noindent
\textbf{Depression Tweet.} This dataset consists of tweets annotated as either depressive or non-depressive. Binary labels are employed as discrete signals for evaluation. We use only the training set from the original dataset and perform our own train/test split to ensure consistent evaluation methodology across all datasets. This dataset was selected based on its use as a benchmark in MentalHelp~\cite{raihan2024mentalhelp}.

\begin{table}[ht]
\centering
\caption{Dataset Statistics}
\label{tab:dataset_stats}
\footnotesize
\resizebox{0.85\linewidth}{!}{%
\begin{tabular}{cccc}
\toprule
\textbf{Dataset} & \textbf{Split} & \textbf{Count} & \textbf{Labels} \\
\midrule
\multirow{2}{*}{\shortstack{Amazon Reviews\\(Software)}} & Train & 10,224 & - \\
                                          & Test  & 2,561 & - \\
\midrule
\multirow{2}{*}{RoSE XSum} & Train & 3,200 & - \\
                                          & Test  & 800 & - \\
\midrule
\multirow{2}{*}{Depression Tweet} & Train & 22,090 & 0: 12,523 / 1: 9,567 \\
                                 & Test  & 5,523 & 0: 3,132 / 1: 2,391 \\
\bottomrule
\end{tabular}
}
\end{table}


\subsection{Experimental Setup}
\label{subsec:setup}

For all experiments, we use Qwen3-1.7B \cite{yang2025qwen3} as our lightweight LLM scorer with a temperature of 0 for deterministic outputs.

\medskip

\subsubsection{Group Partitioning.} For the continuous signal dataset (Amazon), we partition the texts into a $G_{\text{low}}$ and $G_{\text{high}}$ based on the top and bottom percentile of helpful votes. For discrete signal datasets (depression), the groups are naturally defined by their binary labels ($G_{\text{low}}$ for label 0, $G_{\text{high}}$ for label 1), and we sample up to $n=1000$ texts for each group to maintain computational feasibility.

\medskip
    
\subsubsection{Evaluation Criteria.} To ensure a meaningful quality assessment, the semantic criteria scored by the LLM are carefully tailored to the specific context and ground-truth signal of each dataset. This allows us to evaluate qualities that are most relevant to each domain. All criterion-specific prompts were systematically designed with the assistance of GPT-5 as a prompt engineering tool. This process incorporated our evaluation objectives, criterion definitions in \ref{sec:scoring}, and optimizations tailored to the 1.7B model.
\medskip

\noindent
\textbf{Amazon Reviews.} For identifying helpful product reviews, we define five quality criteria based on established review analysis research~\cite{de2014finding}. 
\begin{itemize}[leftmargin=10pt, itemsep=2pt]
    \item \textbf{polarity}: clarity of sentiment expression
    \item \textbf{expertise}: author's demonstrated product knowledge  
    \item \textbf{specificity}: presence of concrete details and examples
    \item \textbf{consistency}: absence of internal contradictions
    \item \textbf{word\_count}: The number of words in the text, scaled according to Eq.~\eqref{eq:scale} with $\sigma_{\text{scale}} = 2$.
\end{itemize}
\medskip

\noindent
\textbf{RoSE XSum Dataset.} We employ criteria grounded in automatic summarization evaluation to assess established quality dimensions of generated summaries~\cite{fabbri2021summeval}.
\begin{itemize}[leftmargin=10pt, itemsep=2pt]
    \item \textbf{coherence}: logical flow and connection between sentences
    \item \textbf{fluency}: grammatical correctness and natural flow of language
    \item \textbf{relevance}: coverage of key points without unrelated information
    \item \textbf{word\_count}: The number of words in the text, scaled according to Eq.~\eqref{eq:scale} with $\sigma_{\text{scale}} = 2$.
\end{itemize}
We exclude consistency as the ACU metric focuses on information inclusion rather than consistency, overlapping with relevance.
\medskip

\noindent
\textbf{Depression Tweet Dataset.} We employ criteria grounded in computational psychology to identify established linguistic markers of depression~\cite{resnik2020discovering, guntuku2019understanding}.
\begin{itemize}[leftmargin=10pt, itemsep=2pt]
    \item \textbf{negative\_affect}: presence of negative emotional language
    \item \textbf{self\_focus}: first-person, self-referential frequency
    \item \textbf{absolutist\_thinking}: extreme, black-and-white language use
    \item \textbf{social\_isolation}: indicators of withdrawal and disconnection
\end{itemize}

\subsection{Baselines and Metrics}
\label{subsec:baselines_metrics}

To rigorously evaluate UniScore, we compare it against several baselines designed to isolate the benefits of our proposed weighting and aggregation method.

\subsubsection{Baselines}
The baselines are chosen to represent simpler or alternative methods for aggregating multi-criteria scores. This comparison is crucial to demonstrate that UniScore's sophisticated approach provides a tangible advantage over standard methods.
\noindent
\begin{itemize}[leftmargin=10pt, itemsep=2pt]
    \item \textbf{Single LLM}: Uses one overall criterion score as final score.
    \item \textbf{Random Weights}: Random weights sampled from $U[-1, 1]$.
    \item \textbf{Linear Regression}: Trained on criteria scores to predict ground-truth signal.
    \item \textbf{Random Forest}: Ensemble regressor predicting ground-truth from criteria scores.
    \item \textbf{Neural Network}: Two-layer MLP (64, 32 units) with ReLU, dropout (0.1), trained with AdamW for up to 100 epochs with early stopping.
\end{itemize}

\subsubsection{Evaluation Metrics}
We employ four metrics to provide a comprehensive view of each method's performance in terms of prediction, discrimination, and consistency.

\begin{itemize}[leftmargin=10pt, itemsep=3pt]
   \item \textbf{Predictive Power}: Measures correlation with ground-truth scores through linear (\textit{Pearson r}), monotonic (\textit{Spearman $\rho$}), and rank-order (\textit{Kendall $\tau$}) metrics.
   \item \textbf{Discriminative Power}: Assesses the ability to distinguish $G_{\text{low}}$ from $G_{\text{high}}$ using Welch's t-test p-values and binary classification metrics (F1-score and accuracy).
   \item \textbf{Consistency}: Examines score stability through coefficient of variation (CV) as a normalized measure of variability and skewness to assess distributional balance, with values closer to zero indicating better symmetry.
\end{itemize}

\subsection{Main Results}
\label{subsec:main_results}

Our experimental results, summarized in Table \ref{tab:main_results}, demonstrate that UniScore consistently and significantly outperforms all baselines across all datasets and metrics.

\begin{table*}[h!]
\centering
\caption{Main performance comparison across all datasets. This indicates UniScore's superior predictive power and its ability to generate consistent, reliable scores. Best results are in \textbf{bold}.}
\label{tab:main_results}
\footnotesize
\resizebox{0.9\textwidth}{!}{%
\begin{tabular}{llccc|ccc|cc}
\toprule
{\textbf{Dataset}} & {\textbf{Method}} & \textbf{Spearman $\rho$} & \textbf{Kendall $\tau$} & Pearson r & p-value & \textbf{F1-score} & Accuracy & CV & Skewness \\
\midrule
{Amazon Reviews}
& Single LLM& -0.1094 & -0.0924 & -0.0427 & 0.0022 & - & - & 0.3628 & -0.4686 \\
& Random Weight& 0.1029 & 0.0784 & 0.0173 & 0.0147 & - & - & 0.6195 & -0.1729 \\
& Regression& 0.3759 & 0.2908 & \textbf{0.2153} & 2.31e-20 & - & - & 1.5904 & 1.6109 \\
& Random Forest& 0.3788 & 0.2952 & 0.0447 & 2.27e-08 & - & - & 5.1330 & 19.1878 \\
& NN& 0.4079 & 0.3158 & 0.2118 & 3.35e-22 & - & - & 1.4832 & 1.6206 \\
& \textbf{UniScore (Ours)} & {} & {} & {} & {} & {} & {} & {} & {} \\
& \textbf{\hspace*{3mm} -- 5\%} & 0.4274 & 0.3308 & 0.1839 & 1.19e-36 & - & - & 0.4001 & -0.0287 \\
& \textbf{\hspace*{3mm} -- 3\%} & 0.4239 & 0.3282 & 0.1856 & \textbf{9.23e-37} & - & - & 0.5363 & \textbf{0.0100} \\
& \textbf{\hspace*{3mm} -- 1\%} & \textbf{0.4303} & \textbf{0.3332} & 0.1936 & 1.89e-33 & - & - & 0.3391 & 0.2478 \\
\midrule
{RoSE XSum}
& Single LLM& 0.0163 & 0.0140 & -0.0209 & 0.3935 & - & - & 0.1294 & 0.5870 \\
& Random Weight& 0.1214 & 0.0874 & 0.1193 & 0.1497 & - & - & 0.1663 & -0.6254 \\
& Regression& 0.1599 & 0.1161 & 0.1447 & 0.1633 & - & - & 0.1407 & -1.1540 \\
& Random Forest& 0.1160 & 0.0841 & 0.1336 & 0.1234 & - & - & 0.4117 & \textbf{0.1674} \\
& NN& \textbf{0.1784} & \textbf{0.1298} & 0.1569 & 0.1319 & - & - & 0.2118 & -0.5202 \\
& \textbf{UniScore (Ours)} & {} & {} & {} & {} & {} & {} & {} & {} \\
& \textbf{\hspace*{3mm} -- 10\%} & 0.1753 & 0.1268 & 0.1574 & 0.0651 & - & - & 0.1013 & -1.3220 \\
& \textbf{\hspace*{3mm} -- 7\%} & 0.1763 & 0.1270 & 0.1583 & \textbf{0.0405} & - & - & 0.1168 & -0.4451 \\
& \textbf{\hspace*{3mm} -- 5\%} & 0.1767 & 0.1279 & \textbf{0.1586} & 0.0434 & - & - & 0.1190 & -0.3918 \\
\midrule
{Depression Tweet}
& Single LLM& - & - & - & < 1e-40 & 0.6997 & 0.7664 & 0.6116 & 0.9141 \\
& Random Weight& - & - & - & < 1e-40 & 0.6958 & 0.7653 & 0.5187 & 1.1140 \\
& Regression& - & - & - & < 1e-40 & 0.7281 & 0.7835 & 0.6901 & 0.3621 \\
& Random Forest& - & - & - & < 1e-40 & 0.7260 & 0.7863 & 0.7015 & 0.3244 \\
& NN& - & - & - & < 1e-40 & 0.7260 & \textbf{0.7865} & 0.6993 & \textbf{0.3181} \\
& \textbf{UniScore (Ours)} & {} & {} & {} & {} & {} & {} & {} & \\
& \textbf{\hspace*{3mm} -- n=1000} & - & - & - & < 1e-40 & \textbf{0.7347} & 0.7675 & 0.1932 & 0.5137 \\
\bottomrule
\end{tabular}
}
\end{table*}

As summarized in Table \ref{tab:main_results}, UniScore validates the aggregation-centric approach: with the same criterion scores, AHP-based weighting yields the strongest or tied-strongest results without additional learning. On Amazon reviews, UniScore attains the best monotonic alignment with ground truth and stable score distributions (low CV and near-zero skewness), while maintaining competitive Pearson $r$. On RoSE XSum, UniScore achieves criterion aggregation performance similar to complex Neural Network models, with a Kendall $\tau$ of 0.1279 that positions it between G-Eval ($\tau$ = 0.120) and current state-of-the-art methods ($\tau$ = 0.148) \cite{stureborg2024large}. On Depression Tweet, UniScore achieves the top F1 and competitive accuracy. This demonstrates the effectiveness of transparent AHP-based aggregation compared to black-box approaches.
\medskip

\begin{figure}[h!]
    \centering
    \includegraphics[width=0.95\linewidth]{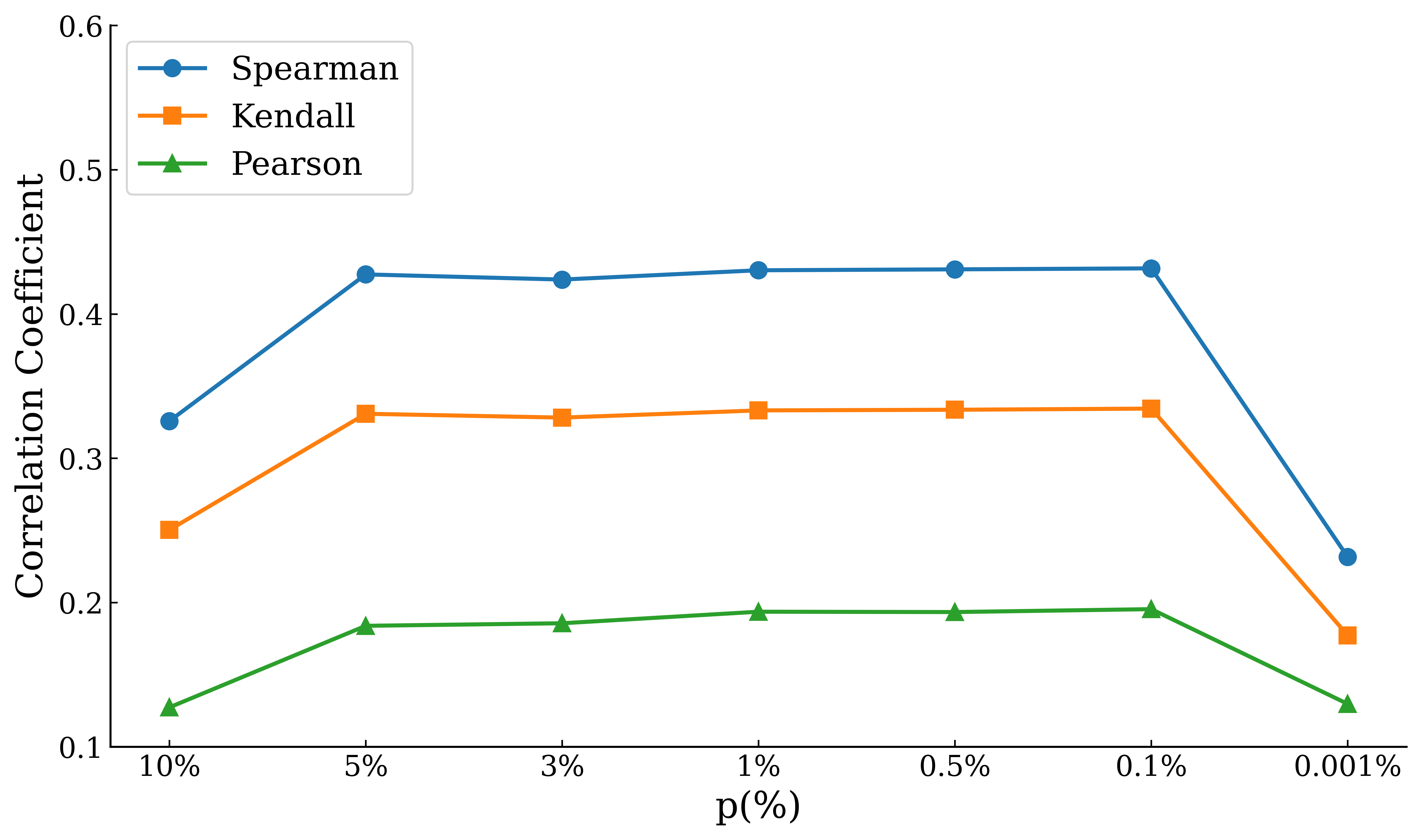}
    \caption{Ablation study of UniScore across $p$ on \textit{Amazon Reviews (Software)}. Correlations remain consistently high for $p \in [0.1\%, 5\%]$, while performance degrades when too strict ($0.001\%$) or too lenient ($10\%$).}
    \label{fig:ablation}
\end{figure}

\textbf{Ablation Study on Hyperparameter p.}
We evaluated UniScore across different $p$s on the Amazon Reviews dataset (\figurename~\ref{fig:ablation}). Performance remains stable and peaks for $p \in [0.1\%, 5\%]$. Overly strict thresholds ($p\leq 0.1\%$) reduce discriminative power by retaining too few samples, while overly lenient thresholds ($p\geq 10\%$) introduce noise from less distinctive examples. Since $p$ determines the number of samples requiring LLM scoring, there is a computational trade-off: smaller $p$ reduces construction time but may sacrifice statistical robustness, while larger $p$ provides more stable estimates at higher cost and reduces discriminative power. This demonstrates the importance of balancing performance and computational efficiency when selecting percentile thresholds.

\subsection{Comparison with Flagship Models}
\label{subsec:flagship}
\begin{figure}[h!]
   \centering
   \includegraphics[width=0.95\linewidth]{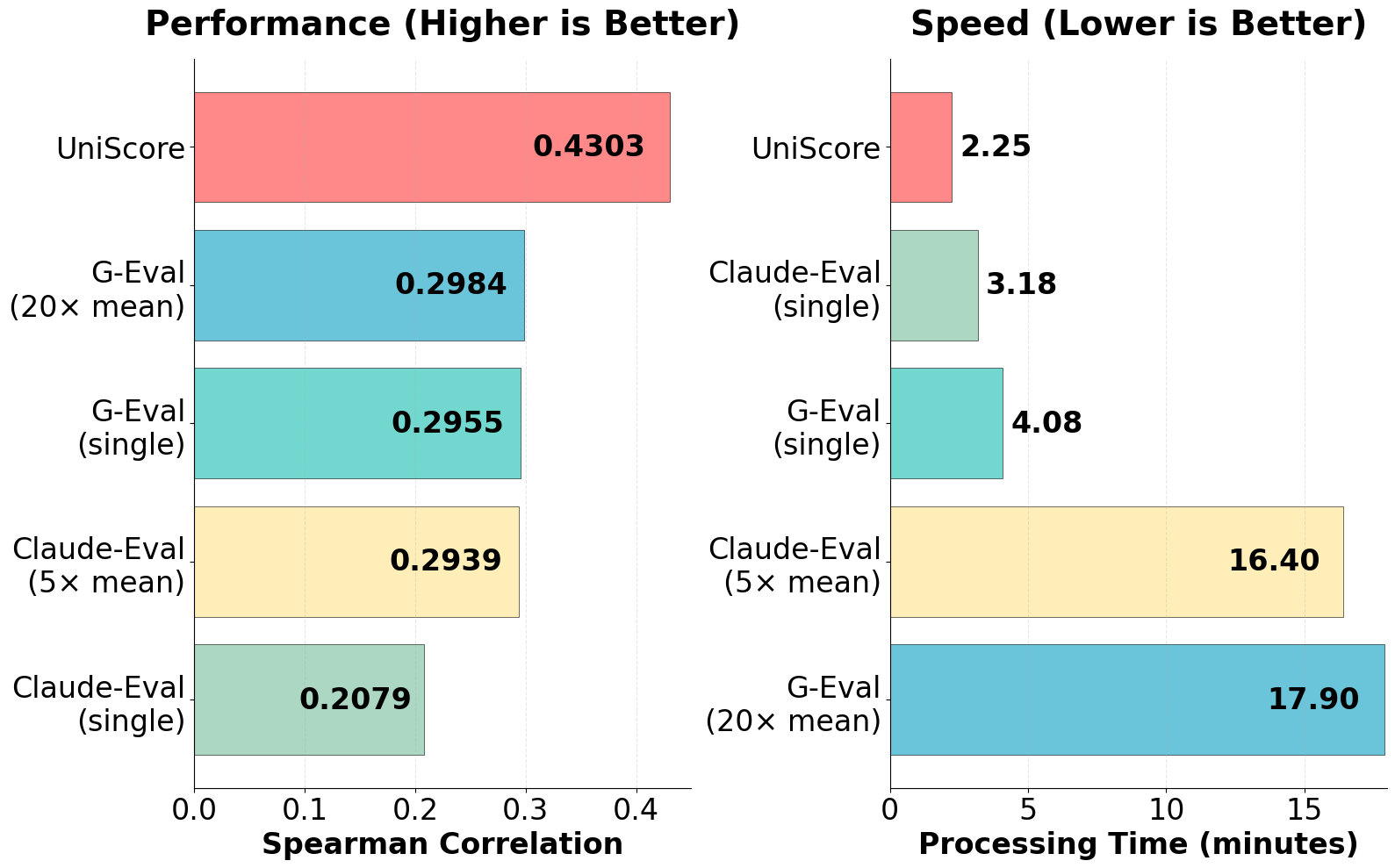}
   \caption{Performance and efficiency comparison of UniScore vs. Flagship model evaluators on \textit{Amazon Reviews}, showing higher correlation and faster processing across all baselines.}
   \label{fig:flagship_comparison}
\end{figure}

Real-world web datasets such as \textit{Amazon} and \textit{Depression} remain largely underexplored in evaluation research, in stark contrast to well-trodden LLM tasks like text summarization. This gap leaves open questions about how evaluation frameworks perform when faced with noisy, user-generated content at scale. To bridge this gap, we benchmark UniScore against state-of-the-art evaluators including G-Eval with GPT-5~\cite{openai2025gpt5} and Claude Sonnet 4~\cite{anthropic2025claude4}, examining the trade-off between efficiency and effectiveness (Figure~\ref{fig:flagship_comparison}). All benchmarks and evaluator prompts were systematically designed with the assistance of the respective models as prompt engineering tools, incorporating our evaluation objectives, criterion definitions in \ref{sec:scoring}.

As shown in Figure \ref{fig:flagship_comparison}, UniScore achieves competitive performance while running locally on a single RTX 3090 GPU. It attains the highest Spearman correlation (0.4303), exceeding both single-run and ensemble baselines. Notably, UniScore completes evaluation of 100 samples in just 2.25 minutes, demonstrating substantial efficiency gains particularly compared to ensemble approaches.

\subsection{Interpretability and Weight Analysis}
\label{subsec:interpretability}
By construction, AHP's mathematical rigor ensures the interpretability of the resulting aggregation weights. Nevertheless, to concretely illustrate the interpretive advantages of our approach, we present a case study on the \textit{Amazon Reviews (Software)} dataset, comparing the final scoring functions derived by UniScore and by a standard linear regression baseline. The two scoring functions differ substantially:
\[
\begin{bmatrix}
\text{UniScore} \\
\text{Regression}
\end{bmatrix} = 
\begin{bmatrix}
-0.11 & 0.17 & 0.18 & -0.07 & 0.47 \\
-0.03 & -0.10 & -0.07 & 0.00 & 0.79
\end{bmatrix}
\begin{bmatrix}
\text{Pol} \\ \text{Exp} \\ \text{Spe} \\ \text{Con} \\ \text{Len}
\end{bmatrix}
\]
where coefficients are normalized to sum to 1 after rounding.

\begin{figure}[h!]
    \centering
    \includegraphics[width=0.95\linewidth]{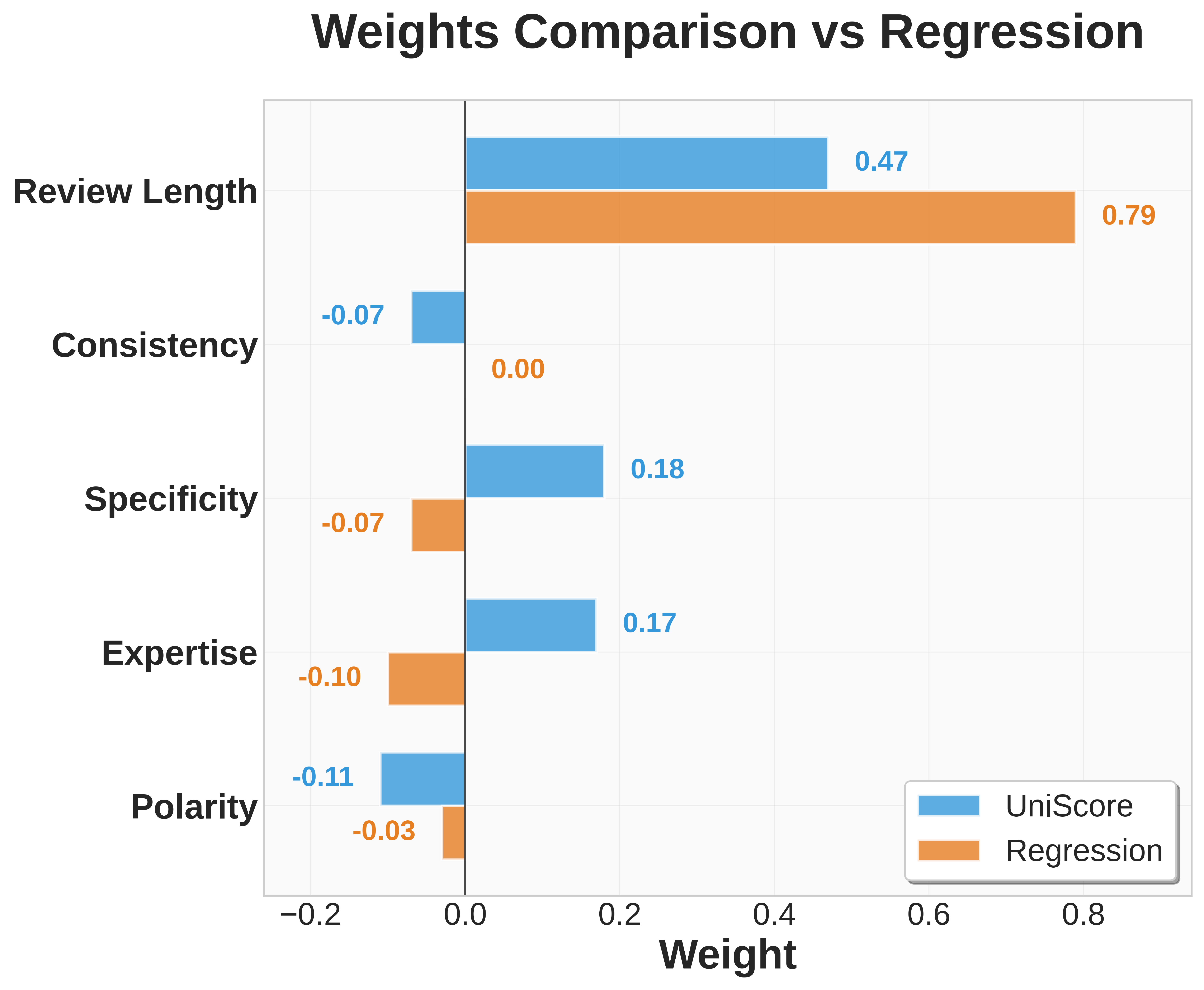}
    \caption{Weight distributions for UniScore and regression on \textit{Amazon Reviews (Software)}.}
    \label{fig:amazon_case_study_weights}
\end{figure}

The regression baseline produces counter-intuitive weights: it assigns negative importance to both \textit{Expertise} and \textit{Specificity}, and overwhelmingly relies on \textit{Review Length} (79\%), effectively ignoring other textual qualities. This contradicts established domain knowledge, which identify expertise and specificity as positive indicators of review helpfulness \cite{de2014finding}. In contrast, UniScore distributes importance more plausibly across criteria, aligning with prior literature and indicating stronger, domain-consistent interpretability. While the weight for \textit{Consistency} was slightly negative, its magnitude was small (7\%), suggesting it had minimal influence on the overall prediction, as can also be seen in Figure~\ref{fig:amazon_case_study_weights}.

\subsection{Consistency Ratio Check}
\label{sec:consistency_exp}
To verify the consistency of the method discussed in Section \ref{sec:ahp}, we performed a Consistency Ratio (CR) check by conducting 100 random samplings for each $p$ (1\%, 3\%, and 5\%). As shown in Table \ref{tab:cr_check}, not a single sampled CR value exceeded the conventional consistency error threshold of 0.1\cite{Saaty1980}. This result experimentally suggests that our difference-based approach does not have significant consistency issues.

\begin{table}[h!]
    \centering
    \caption{Consistency Ratio (CR) Statistics across Different p\%}
    \begin{tabular}{cccc}
        \toprule
        p-value  & Mean & Variance & Max \\
        \midrule
        5\%         & 0.0123 & 7.88e-05 & 0.0277 \\
        3\%          & 0.0169 & 7.83e-05 & 0.0389 \\
        1\%         & 0.0123 & 3.34e-05 & 0.0378 \\
        \bottomrule
    \end{tabular}
    \label{tab:cr_check}
\end{table}

\section{Discussion}
\label{sec:discussion}

This study demonstrates that UniScore outperforms existing automated evaluation methods in interpretability, computational efficiency, and predictive performance, as detailed in our experimental results (Section \ref{sec:experiments}). These advantages stem from its mathematically grounded design and operational structure.

First, UniScore ensures interpretability from the weight derivation stage. Black-box models are inherently difficult to interpret, while regression models, although structurally interpretable, often produce weights that contradict domain intuition, as observed in our case study (Section \ref{subsec:interpretability}, Figure \ref{fig:amazon_case_study_weights}). In contrast, UniScore applies the AHP using the principal eigenvector method to produce consistent weights. The contribution of each criterion is numerically quantified (Table \ref{tab:main_results}), offering domain experts results that are both intuitive and persuasive.

Second, UniScore achieves high performance with a simple and efficient computational structure. The final score is computed as a linear combination of multi-criteria scores, without requiring large-scale neural networks or complex feature transformations. As demonstrated in our flagship model comparison (Section \ref{subsec:flagship}), UniScore delivers near real-time speed while exceeding the evaluation quality of flagship models (Figure \ref{fig:flagship_comparison}).

Third, UniScore’s performance benefits from its distribution-based design. Instead of relying solely on absolute values of scores, it employs JSD to capture stable and symmetric differences between group distributions. This information-theoretic property makes the framework robust to outliers while preserving discriminative power, supporting consistent performance across both continuous and discrete data, as validated by our main results (Table \ref{tab:main_results}).

In summary, UniScore achieves a unique balance by combining interpretability from mathematically rigorous weighting, speed and resource efficiency from its lightweight structure, and performance consistency backed by the information-theoretic stability of JSD. These characteristics make it a balanced evaluation framework across theoretical, practical, and performance dimensions, enabling high-quality automated text evaluation even in resource-constrained environments and suggesting broad applicability across diverse domains and data types.

\section{Conclusion}
\label{sec:conclusion}

In this paper, we introduced UniScore, a novel framework for automatically generating an interpretable, efficient, and high-performance text quality scoring function. By integrating multi-criteria LLM-based evaluation with AHP-based weighting mechanism, UniScore performs comparably to or surpasses traditional baselines.

The practical implications of UniScore are substantial, particularly for real-time web services and industrial applications. Its low latency enables on-the-fly evaluation of user content across domains like e-commerce reviews, digital mental healthcare, and automated essay scoring (AES), while LLM-generated explanations combined with UniScore's weights provide users with convincing score justifications. It also supports semi-supervised operation, partitioning unlabeled datasets into $G_{\text{high}}$ and $G_{\text{low}}$ according to system configuration for hybrid use with labeled data.

Our framework provides a foundation for future research. Despite its robustness, it remains dependent on a pre-defined signal, meaning incomplete references may constrain performance. Future work should explore achieving robust discriminative power from incomplete data without a perfect reference signal.

Additional limitations include the use of a limited set of LLMs, which may not reflect the full diversity of modern models. The approach also requires user-defined evaluation criteria, requiring human intervention in framework design. Finally, as experiments used lightweight models for real-time web integration, the effectiveness of UniScore with large-scale models remains unknown.

With its balance of efficiency and interpretability, UniScore has the potential to set a new standard for multi-criteria text evaluation in academia and industry. The proposed JSD-AHP method offers a generalizable way to build interpretable linear models from any feature set, with potential applications beyond LLM-generated text, such as tabular data analysis. This work lays a strong foundation for future advances in transparent, reliable, and adaptive automated scoring.


\bibliographystyle{ACM-Reference-Format}
\bibliography{base}

\end{document}